\documentclass[aip,jcp,preprint,groupedaddress]{revtex4-1}%
\usepackage{graphicx}
\usepackage{tabularx}
\usepackage{dcolumn}
\usepackage{longtable}
\usepackage{tensor}
\usepackage{placeins}
\usepackage{bm}
\usepackage{amsmath}
\usepackage{amsfonts}
\usepackage{amssymb}
\usepackage{float}
\usepackage{xcolor}%
\providecommand{\U}[1]{\protect\rule{.1in}{.1in}}
\begin{document}
\title{On-the-fly \textit{ab initio} semiclassical evaluation of vibronic spectra at
finite temperature}
\author{Tomislav Begu\v{s}i\'{c}}
\email{tomislav.begusic@epfl.ch}
\author{Ji\v{r}\'i Van\'i\v{c}ek}
\email{jiri.vanicek@epfl.ch}
\affiliation{Laboratory of Theoretical Physical Chemistry, Institut des Sciences et
Ing\'enierie Chimiques, Ecole Polytechnique F\'ed\'erale de Lausanne (EPFL),
CH-1015, Lausanne, Switzerland}
\date{\today}

\begin{abstract}
To compute and analyze vibrationally resolved electronic spectra at zero
temperature, we have recently implemented the on-the-fly \textit{ab initio}
extended thawed Gaussian approximation [A. Patoz \textit{et al.}, J. Phys.
Chem. Lett. \textbf{9}, 2367 (2018)], which accounts for anharmonicity,
mode-mode coupling, and Herzberg--Teller effects. Here, we generalize this
method in order to evaluate spectra at non-zero temperature. In line with
thermo-field dynamics, we transform the von Neumann evolution of the
coherence component of the density matrix to the Schr\"{o}dinger
evolution of a wavefunction in an augmented space with twice as many degrees
of freedom. Due to the efficiency of the extended thawed Gaussian approximation,
this increase in the number of coordinates results in nearly no additional
computational cost. More specifically, compared to the original,
zero-temperature approach, the finite-temperature method requires no
additional \textit{ab initio} electronic structure calculations. At the same
time, the new approach allows for a clear distinction among
finite-temperature, anharmonicity, and Herzberg--Teller effects on spectra. We
show, on a model Morse system, the advantages of the finite-temperature thawed
Gaussian approximation over the commonly used global harmonic methods and
apply it to evaluate the symmetry-forbidden absorption spectrum of benzene,
where all of the aforementioned effects contribute.

\end{abstract}
\maketitle

\graphicspath{{"C:/Users/GROUP LCPT/Documents/Group/Tomislav/FiniteTemperature/figures/"}
{./figures/}{C:/Users/Jiri/Dropbox/Papers/Chemistry_papers/2020/FiniteTemperature/figures/}}

\section{\label{sec:intro}Introduction}

Vibrationally resolved electronic spectra have, for a very long time, been
used to learn more about electronic and vibrational states of molecules, their
potential energy surfaces, and light-induced dynamics of
nuclei.\cite{Herzberg:1966,Hollas:2004,Quack_Merkt:2011,book_Heller:2018} The
computational methods for simulating such spectra are, therefore, an essential
tool in physical chemistry.

The most widespread is the global harmonic method,\cite{Santoro_Barone:2007,
Santoro_Barone:2008,Barone_Santoro:2009} which employs the harmonic
approximation for both ground- and excited-state potential energy surfaces.
Within the framework of the global harmonic approximation, one can easily
account for non-Condon and finite-temperature
effects.\cite{Niu_Shuai:2010,Borrelli_Peluso:2012,Baiardi_Barone:2013,Reddy_Prasad:2015,Reddy_Prasad:2016}
This approximation, however, neglects the effects of anharmonicity, which can
significantly alter molecular spectra. Other
quantum\cite{Ben-Nun_Martinez:1999,Meng_Meyer:2013,Bonfanti_Pollak:2018,Picconi_Burghardt:2019}
and
semiclassical\cite{Hwang_Warshel:1985,Mukamel_Yan:1989,Tatchen_Pollak:2009,Zimmermann_Vanicek:2014,Bonfanti_Pollak:2018,book_Heller:2018}
methods do include anharmonicity effects on spectra, but at a substantial
computational cost. Recently, we have been investigating the thawed Gaussian
approximation,\cite{Heller:1975, Wehrle_Vanicek:2014,
Wehrle_Vanicek:2015,Begusic_Vanicek:2019} an efficient semiclassical method
that accounts partially for anharmonicity and requires no initial knowledge
of the potential energy surface. The method has been extended to include
non-Condon effects, namely, to account for the Herzberg--Teller contribution
to the transition dipole
moment.\cite{Patoz_Vanicek:2018,Begusic_Vanicek:2018,Prlj_Vanicek:2020}
Unfortunately, as a wavepacket propagation method, it has been limited to
computing spectra in the zero-temperature limit, where only the ground
vibrational state is populated initially.

To account for non-zero temperature, one typically employs the density matrix
formalism, where a number of numerically
exact\cite{Tanimura_Kubo:1989,Tang_Wu:2015,Chen_Tanimura:2015} and
approximate\cite{Mukamel:1982,Bergsma_Heller:1984,book_Mukamel:1999,Crespo-Otero_Barbatti:2012,Vanicek:2017,Kossoski_Barbatti:2018,Karsten_Kuhn:2018}
approaches exist. Otherwise, typical wavefunction-based methods can be used in
combination with statistical sampling of initial
conditions.\cite{Manthe_Larranaga:2001,Gelman_Kosloff:2003,Wang_Thoss:2006,Nest_Kosloff:2007,Meyer_Worth:2009,Lorenz_Saalfrank:2014,Wang_Zhao:2017}
Thermo-field dynamics\cite{Suzuki:1985,Takahashi_Umezawa:1996} offers an
alternative way to make wavefunction-based methods applicable at finite
temperature: the problem, which seemingly requires the von Neumann equation
for the density matrix, is mapped to a time-dependent Schr\"{o}dinger equation
with twice as many degrees of freedom. Recently, the thermo-field dynamics was
employed in chemistry for solving the coupled electronic-vibrational
dynamics,\cite{Borrelli_Gelin:2016,Borrelli_Gelin:2017,Gelin_Borrelli:2017,Chen_Zhao:2017}
electronic structure,\cite{Harsha_Scuseria:2019} and vibronic
spectroscopy\cite{Reddy_Prasad:2015} problems. The application to vibronic
spectroscopy, which is of central interest to this work, was, however,
restricted to the global harmonic approximation.

Here, we combine the extended thawed Gaussian wavepacket propagation with the
thermo-field dynamics in order to include both anharmonicity and
finite-temperature effects. Due to the favorable scaling of the thawed
Gaussian approximation with the system's size, the new method
adds nearly no additional cost to the original, zero-temperature approach. To
illustrate the accuracy achieved by going beyond both global harmonic and
zero-temperature approximations, we test the method on a set of Morse
potentials with different degrees of anharmonicity and at different
temperatures. Finally, we apply it to evaluate the spectrum corresponding to
the symmetry-forbidden electronic transition $\text{S}_{1}\leftarrow
\text{S}_{0}$ ($\tilde{\text{A}}^{1}\text{B}_{\text{2u}}\leftarrow
\tilde{\text{X}}^{1}\text{A}_{\text{1g}}$) of benzene and demonstrate that the
simultaneous inclusion of Herzberg--Teller, anharmonicity, and
finite-temperature effects is needed to reproduce the experimental spectrum.

\section{\label{sec:theory}Theory}

\subsection{\label{subsec:tga_0}Extended thawed Gaussian approximation for
zero-temperature spectra}

Before turning to vibrationally resolved electronic spectra at finite
temperature, let us briefly describe the original, zero-temperature approach
based on the extended thawed Gaussian approximation.

The absorption spectrum at zero temperature can be computed as the Fourier
transform,\cite{Heller:1981a,book_Mukamel:1999,book_Tannor:2007}
\begin{equation}
\sigma(\omega)=\frac{4\pi\omega}{\hbar c}\text{Re}\int_{0}^{\infty
}dtC(t)e^{i\omega t}, \label{eq:sigma}%
\end{equation}
of the correlation function,
\begin{equation}
C(t)=\langle1,\text{g}|\hat{\mu}^{\dagger}e^{-i\hat{H}_{2}t/\hbar}\hat{\mu
}|1,\text{g}\rangle e^{i\omega_{1,\text{g}}t}, \label{eq:C_t_0K}%
\end{equation}
where $|1,\text{g}\rangle$ is the ground (\textquotedblleft
g\textquotedblright) vibrational state of the ground (\textquotedblleft%
1\textquotedblright) electronic state, $\hbar\omega_{1,\text{g}}$ is its
energy, $\hat{H}_{2}$ is the nuclear Hamiltonian corresponding to the excited
(subscript \textquotedblleft2\textquotedblright) electronic state, and
$\hat{\mu}$ is the transition dipole moment $\hat{\mu}_{21}=\hat{\vec{\mu}%
}_{21}\cdot\vec{\epsilon}$ projected on the polarization $\vec{\epsilon}$ of
the external electric field. In other words, to compute the spectrum, one has
to evolve the nuclear wavefunction $|\phi_{0}\rangle=\hat{\mu}|1,\text{g}%
\rangle$ on the excited-state surface, which is, in general, a challenging
task that scales exponentially with the number of atoms.

Different exact
quantum\cite{Burghardt_Giri:2008,Meyer_Worth:2009,Saita_Shalashilin:2012,Richings_Lasorne:2015,Curchod_Martinez:2018}
and
semiclassical\cite{Herman_Kluk:1984,Miller:2001,Grossmann:2006,Tatchen_Pollak:2009,Ceotto_Atahan:2009,Ceotto_Atahan:2009a,Ceotto_Aspuru-Guzik:2011,Wong_Roy:2011,Ianconescu_Pollak:2013,Buchholz_Ceotto:2016,DiLiberto_Ceotto:2016,Buchholz_Ceotto:2017,Gabas_Ceotto:2017,Buchholz_Ceotto:2018,Gabas_Ceotto:2018,Conte_Ceotto:2019,Gabas_Ceotto:2019,Micciarelli_Ceotto:2019}
methods were developed for solving the problem of wavepacket propagation.
Sometimes, the region of the excited-state potential energy surface explored
by the evolved wavepacket is fairly harmonic, meaning that it can be well
approximated by a second-order Taylor expansion in nuclear coordinates about a
reference geometry; we call this the \textit{global harmonic approximation}.
Then, the correlation function (\ref{eq:C_t_0K}) can be obtained
analytically.\cite{Niu_Shuai:2010,Borrelli_Peluso:2012,Baiardi_Barone:2013,Tapavicza_Sundholm:2016,
Tapavicza:2019} Moreover, in this case the excited-state surface is easily
constructed from a single Hessian calculation. The prevalence of
the global harmonic method in the vibronic spectroscopy
literature\cite{Dierksen_Grimme:2004,
Biczysko_Barone:2009,Barone_Santoro:2009,Niu_Shuai:2010,Baiardi_Barone:2013,Tapavicza_Sundholm:2016,Fortino_Pedone:2019,Tapavicza:2019}
testifies, on the one hand, to its applicability in a wide range of molecules
and, on the other hand, to the absence of accessible alternatives that can
account for anharmonicity effects. Often, due to the reduced resolution of
electronic spectra, even such a crude approximation, which would nowadays be
almost unacceptable for the simulation of vibrational (infrared) spectra, is
considered appropriate.

To account for the anharmonicity effects on the spectrum at least
approximately, we recommend using the simple and efficient semiclassical
\textit{thawed Gaussian approximation}.\cite{Heller:1975} In contrast to many
other exact or approximate quantum dynamics methods, this method is
computationally feasible even for rather large molecules and can be employed
in a \textquotedblleft black-box\textquotedblright\ fashion, i.e., it requires
little human input. In particular, the thawed Gaussian approximation requires
only local potential energy information along the classical trajectory (as
described below) and, therefore, supports an on-the-fly implementation where
the potential energy is provided by an \textit{ab initio} electronic structure calculation.

Within the thawed Gaussian approximation,\cite{Heller:1975} the wavepacket is
assumed to be a complex Gaussian function
\begin{equation}
\psi_{t}(q)=e^{\frac{i}{\hbar}\left[  \frac{1}{2}(q-q_{t})^{T}\cdot A_{t}%
\cdot(q-q_{t})+p^{T}\cdot(q-q_{t})+\gamma_{t}\right]  } \label{eq:tga_wp}%
\end{equation}
parametrized by the time-dependent $D$-dimensional real vectors $q_{t}$ and
$p_{t}$, $D\times D$ complex symmetric matrix $A_{t}$, and complex number
$\gamma_{t}$; $D$ is the number of degrees of freedom. The time dependence of
the matrix $A_{t}$ implies that the width of the thawed Gaussian wavepacket
changes with time, as opposed to the \textit{frozen} Gaussian ansatz where the
width remains constant. Wavepacket (\ref{eq:tga_wp}) solves exactly the
Schr\"{o}dinger equation
\begin{equation}
i\hbar|\dot{\psi}_{t}\rangle=[T(\hat{p})+V_{\text{LHA}}(\hat{q})]|\psi
_{t}\rangle, \label{eq:Sch}%
\end{equation}
where $T(p)=\frac{1}{2}p^{T}\cdot m^{-1}\cdot p$ is the kinetic energy and
\begin{equation}
V_{\text{LHA}}(q)=V(q_{t})+V^{\prime}(q_{t})^{T}\cdot(q-q_{t})+\frac{1}%
{2}(q-q_{t})^{T}\cdot V^{\prime\prime}(q_{t})\cdot(q-q_{t}) \label{eq:V_LHA}%
\end{equation}
is the \textit{local harmonic approximation} of the true potential energy
$V(q)$ about $q_{t}$, if the time-dependent parameters of $\psi_{t}$ satisfy
the following equations of motion:\cite{Heller:1975}
\begin{align}
\dot{q}_{t}  &  =m^{-1}\cdot p_{t},\label{eq:q_t_dot}\\
\dot{p}_{t}  &  =-V^{\prime}(q_{t}),\label{eq:p_t_dot}\\
\dot{A}_{t}  &  =-A_{t}\cdot m^{-1}\cdot A_{t}-V^{\prime\prime}(q_{t}%
),\label{eq:A_t_dot}\\
\dot{\gamma}_{t}  &  =L_{t}+\frac{i\hbar}{2}\text{Tr}(m^{-1}\cdot A_{t}).
\label{eq:gamma_t_dot}%
\end{align}
In these equations, $V^{\prime}(q_{t})$ and
$V^{\prime\prime}(q_{t})$ denote, respectively, the gradient and Hessian of the potential energy evaluated at
$q_{t}$, $m$ is the symmetric mass matrix, and $L_{t}=T(p_{t})-V(q_{t})$ is
the Lagrangian. Note that due to the local harmonic
approximation, Eq.~(\ref{eq:Sch}) is a \emph{nonlinear} Schr\"{o}dinger
equation because the potential $V_{\text{LHA}}(q)$ depends on the wavefunction
through the parameter $q_{t}$ [Eq.~(\ref{eq:V_LHA})], i.e., $V_{\text{LHA}%
}(q)\equiv V_{\text{LHA}}(q;q_{t})\equiv V_{\text{LHA}}(q;\psi_{t})$.

To construct the initial wavepacket, the ground-state potential energy surface
$V_{1}(q)$ is assumed to be harmonic in the vicinity of its minimum
$q_{\text{eq}}$, i.e., the ground-state Hamiltonian is approximated as
\begin{equation}
H_{1}(q)\approx-\frac{\hbar^{2}}{2}\partial_{q}^{T}\cdot m^{-1}\cdot
\partial_{q}+\frac{1}{2}(q-q_{\text{eq}})^{T}\cdot K\cdot(q-q_{\text{eq}}),
\label{eq:H_1_HA}%
\end{equation}
where $K=V_{1}^{\prime\prime}(q_{\text{eq}})$ is the symmetric force-constant
matrix and $\partial_{q}=\partial/\partial q$. In position representation, the
lowest eigenstate $\psi_{0}(q)$ of the Hamiltonian (\ref{eq:H_1_HA}) is a
Gaussian (\ref{eq:tga_wp}) with parameters
\begin{align}
q_{0}  &  =q_{\text{eq}},\label{eq:q_0}\\
p_{0}  &  =0,\label{eq:p_0}\\
A_{0}  &  =im^{1/2}\cdot\Omega\cdot m^{1/2},\label{eq:A_0}\\
\gamma_{0}  &  =(-i\hbar/4)\ln[\det(\operatorname{Im}A_{0}/\pi\hbar)],
\label{eq:gamma_0}%
\end{align}
where $\Omega=\sqrt{m^{-1/2}\cdot K\cdot m^{-1/2}}$. This initial wavefunction
$\psi_{0}(q)$ is then evolved by solving differential equations
(\ref{eq:q_t_dot})--(\ref{eq:gamma_t_dot}) with $V=V_{2}$ (the excited-state
potential energy).

The thawed Gaussian wavepacket (\ref{eq:tga_wp}) is not suited to treat
non-Condon effects, i.e., the effects due to the dependence of the transition
dipole moment $\mu(q)$ on nuclear coordinates $q$. Within the Herzberg--Teller
approximation---the simplest extension of the Condon approximation---the
transition dipole moment is assumed to be a linear
function\cite{Herzberg_Teller:1933}
\begin{equation}
\mu(q)=\mu(q_{\text{eq}})+\mu^{\prime}(q_{\text{eq}})^{T}\cdot(q-q_{\text{eq}%
}), \label{eq:mu_HT}%
\end{equation}
where $\mu^{\prime}(q_{\text{eq}})$ is the gradient of $\mu$ with respect to
nuclear coordinates at the equilibrium geometry. Then, $\phi_{0}(q)=\mu
(q)\psi_{0}(q)$ is no longer a Gaussian wavepacket. Fortunately, the
\textit{extended} thawed Gaussian
ansatz,\cite{Lee_Heller:1982,Patoz_Vanicek:2018}
\begin{equation}
\phi_{t}(q)=[a_{t}+b_{t}^{T}\cdot(q-q_{t})]\psi_{t}(q), \label{eq:etga}%
\end{equation}
which is a special case of Hagedorn's \textquotedblleft Gaussian times a
polynomial\textquotedblright%
\ wavepacket,\cite{Hagedorn:1998,Faou_Lubich:2009,Lasser_Lubich:2020} solves
the same Schr\"{o}dinger equation [Eq.~(\ref{eq:Sch})] as $\psi_{t}(q)$,
provided that the Gaussian parameters evolve, as before, according to
Eqs.~(\ref{eq:q_t_dot})--(\ref{eq:gamma_t_dot}) and, in addition,
\begin{align}
\dot{a}_{t}  &  =0,\label{eq:a_t}\\
\dot{b}_{t}  &  =-A_{t}\cdot m^{-1}\cdot b_{t}. \label{eq:b_t_dot}%
\end{align}
Hence, with the extended thawed Gaussian approximation, one can include the
Herzberg--Teller contribution at nearly no additional computational cost.

\subsection{\label{subsec:spectraT}Vibrationally resolved electronic spectra
at finite temperature}

At non-zero temperature, the dipole-dipole correlation function
needed in vibronic spectroscopy is
\begin{equation}
C(t)=\text{Tr}(\hat{\mu}^{\dagger}e^{-i\hat{H}_{2}t/\hbar}\hat{\mu}\hat{\rho
}e^{i\hat{H}_{1}t/\hbar}), \label{eq:C_t_1}%
\end{equation}
where $\hat{\rho}=e^{-\beta\hat{H}_{1}}/\text{Tr}(e^{-\beta\hat{H}_{1}})$ is
the vibrational density operator and $\beta=1/k_{B}T$. Note that in
Eq.~(\ref{eq:C_t_1}) we assumed that only the ground electronic state is
populated in the thermal equilibrium, which is usually justified by the large
energy gap between the ground and first excited electronic states.
Because the time evolution in Eq.~(\ref{eq:C_t_1}) involves two
different Hamiltonians, an obvious classical analogue of Eq.~(\ref{eq:C_t_1})
is missing, which, in turn, hinders the development of classical-like or
semiclassical approximations for $C(t)$. Here, we demonstrate that by
transforming the problem to the one of wavepacket dynamics in an augmented
space one can easily make use of the existing semiclassical methods for
solving the time-dependent Schr\"{o}dinger equation.

The correlation function can be re-written as
\begin{align}
C(t)  &  =\text{Tr}(\hat{\rho}^{1/2}\hat{\mu}^{\dagger}e^{-i\hat{H}_{2}%
t/\hbar}\hat{\mu}\hat{\rho}^{1/2}e^{i\hat{H}_{1}t/\hbar})\label{eq:C_t_2}\\
&  =\int dqdq^{\prime}\langle q^{\prime}|\hat{\rho}^{1/2}\hat{\mu}^{\dagger
}|q\rangle\langle q|e^{-i\hat{H}_{2}t/\hbar}\hat{\mu}\hat{\rho}^{1/2}%
e^{i\hat{H}_{1}t/\hbar}|q^{\prime}\rangle\label{eq:C_t_3}\\
&  =\int d\bar{q}\bar{\phi}_{0}(\bar{q})^{\ast}\bar{\phi}_{t}(\bar{q}),
\label{eq:C_t_7}%
\end{align}
where
\begin{align}
\bar{\phi}_{0}(\bar{q})  &  =\langle q|\hat{\mu}\hat{\rho}^{1/2}|q^{\prime
}\rangle,\label{eq:phi_0_bar}\\
\bar{\phi}_{t}(\bar{q})  &  =\langle q|e^{-i\hat{H}_{2}t/\hbar}\hat{\mu}%
\hat{\rho}^{1/2}e^{i\hat{H}_{1}t/\hbar}|q^{\prime}\rangle\\
&  =e^{-iH_{2}(q)t/\hbar}e^{iH_{1}(q^{\prime})t/\hbar}\langle q|\hat{\mu}%
\hat{\rho}^{1/2}|q^{\prime}\rangle\label{eq:C_t_4}\\
&  =e^{-i[H_{2}(q)-H_{1}(q^{\prime})]t/\hbar}\langle q|\hat{\mu}\hat{\rho
}^{1/2}|q^{\prime}\rangle\label{eq:C_t_5}\\
&  =e^{-i\bar{H}(\bar{q})t/\hbar}\bar{\phi}_{0}(\bar{q}), \label{eq:C_t_6}%
\end{align}
$\bar{q}=(q,q^{\prime})^{T}$ is a $2D$-dimensional coordinate vector, and
\begin{equation}
\bar{H}(\bar{q})=H_{2}(q)-H_{1}(q^{\prime})
\end{equation}
is a Hamiltonian in $\bar{q}$ coordinates. In Eq.~(\ref{eq:C_t_2}), we used the
relation $[\hat{\rho},\hat{H}_{1}]=0$ and the cyclic property of the trace; in
Eq.~(\ref{eq:C_t_3}), we introduced the position representation in $q$ and
$q^{\prime}$ coordinates; in going from (\ref{eq:C_t_4}) to (\ref{eq:C_t_5}),
we used the fact that the two Hamiltonians $H_{1}(q^{\prime})$ and $H_{2}(q)$
commute because they act on different coordinates; finally,
Eq.~(\ref{eq:C_t_7}) follows from (\ref{eq:C_t_3}) because
\begin{equation}
\langle q^{\prime}|\hat{\rho}^{1/2}\hat{\mu}^{\dagger}|q\rangle=\langle
q|\hat{\mu}\hat{\rho}^{1/2}|q^{\prime}\rangle^{\ast}=\bar{\phi}_{0}(\bar
{q})^{\ast}%
\end{equation}
and $(\hat{\rho}^{1/2})^{\dagger}=\hat{\rho}^{1/2}$.

Equation~(\ref{eq:C_t_7}) has a remarkable interpretation---the dipole-dipole
correlation function $C(t)$ for a $D$-dimensional system at finite temperature
$T$ can be thought of as a wavepacket autocorrelation function of $\bar{\phi
}_{t}(\bar{q})$ evolved with the Hamiltonian $\bar{H}(\bar{q})$ according to
the Schr\"{o}dinger equation,
\begin{equation}
i\hbar\dot{\bar{\phi}}_{t}(\bar{q})=\bar{H}(\bar{q})\bar{\phi}_{t}(\bar{q}),
\label{eq:Sch_T}%
\end{equation}
which describes an effective $2D$-dimensional system at zero temperature.

The approach described here is, despite the explicit use of the position
representation, equivalent to the thermo-field dynamics, as presented in
Ref.~\onlinecite{Reddy_Prasad:2015}. Indeed, the final result does not depend
on the representation:
\begin{equation}
C(t)=\int d\bar{q}\bar{\phi}_{0}(\bar{q})^{\ast}\bar{\phi}_{t}(\bar{q})=\int
d\bar{q}\langle\bar{\phi}_{0}|\bar{q}\rangle\langle\bar{q}|\bar{\phi}%
_{t}\rangle=\langle\bar{\phi}_{0}|\bar{\phi}_{t}\rangle, \label{eq:C_t_norep}%
\end{equation}
where $|\bar{q}\rangle=|q\rangle|\tilde{q}^{\prime}\rangle$ is a general
position state in the augmented direct-product Hilbert space and $|\tilde
{q}\rangle$ denotes a position state in the \textquotedblleft
fictitious\textquotedblright\ (or \textquotedblleft tilde\textquotedblright)
Hilbert space. In Appendix~\ref{app_sec:TFD}, we derive
Eq.~(\ref{eq:C_t_norep}) using standard thermo-field dynamics notation and
without invoking the position representation.

In principle, any known method for solving the time-dependent Schr\"odinger
equation can be applied to obtain $\bar{\phi}_{t}$. However, the doubled
number of coordinates adds a substantial, if not prohibitive, computational
cost to the already large cost of zero-temperature calculations with
exponentially-scaling exact quantum methods. In the following, we therefore
employ the extended thawed Gaussian approximation, which scales favorably with
the number of degrees of freedom.

\subsection{\label{subsec:tga_T}Extended thawed Gaussian approximation for
finite-temperature spectra}

To solve Eq.~(\ref{eq:Sch_T}) with the (extended) thawed Gaussian
approximation, we must first identify $\bar{\phi}_{0}$ and the local harmonic
approximation to the potential energy $\bar{V}(\bar{q}) = V_{2}(q) -
V_{1}(q^{\prime})$.

If we assume, as in Sec.~\ref{subsec:tga_0}, that the ground-state surface
$V_{1}$ is harmonic [Eq.~(\ref{eq:H_1_HA})], a general off-diagonal matrix
element $\rho^{1/2}(q,q^{\prime})\equiv\rho^{1/2}(\bar{q})$ of $\hat{\rho
}^{1/2}$ is a Gaussian (\ref{eq:tga_wp}) parametrized with $2D$-dimensional
vectors
\[
\bar{q}_{0}=%
\begin{pmatrix}
q_{\text{eq}}\\
q_{\text{eq}}%
\end{pmatrix}
,\qquad\bar{p}_{0}=%
\begin{pmatrix}
0\\
0
\end{pmatrix}
,
\]
a $2D\times2D$ matrix
\begin{equation}
\bar{A}_{0}=i%
\begin{pmatrix}
\text{A} & \text{B}\\
\text{B} & \text{A}%
\end{pmatrix}
\label{eq:A_0_temp}%
\end{equation}
composed of $D\times D$ submatrices
\begin{align}
\text{A}  &  =m^{1/2}\cdot\Omega\cdot\coth(\beta\hbar\Omega/2)\cdot
m^{1/2},\label{eq:A_0_temp_A}\\
\text{B}  &  =-m^{1/2}\cdot\Omega\cdot\sinh(\beta\hbar\Omega/2)^{-1}\cdot
m^{1/2}, \label{eq:A_0_temp_B}%
\end{align}
and a scalar
\begin{equation}
\bar{\gamma}_{0}=(-i\hbar/2)\ln[\det(m\cdot\Omega/\pi\hbar)].
\label{eq:gamma_0_temp}%
\end{equation}
See Appendix~\ref{app_sec:density} for the derivation of
Eqs.~(\ref{eq:A_0_temp})--(\ref{eq:gamma_0_temp}). The harmonic approximation for
the ground-state potential energy surface is justified in the vicinity of its
minimum and, therefore, for the construction of the equilibrium vibrational
density matrix. In fact, even in fairly anharmonic systems, Gaussian density
matrix often serves as a good starting point for semiclassical
approximations.\cite{Wang_Miller:1998, Liu_Miller:2006, Liu_Miller:2009,
Liu:2014} Next, we assume $\hat{\mu}$ to be diagonal in position
representation,
\begin{equation}
\bar{\phi}_{0}(\bar{q})=\mu(q)\rho^{1/2}(\bar{q}),
\end{equation}
and employ the Herzberg--Teller approximation [Eq.~(\ref{eq:mu_HT})] to
obtain
\begin{align}
\bar{\phi}_{0}(\bar{q})  &  =[\mu(q_{\text{eq}})+\bar{b}_{0}^{T}\cdot(\bar
{q}-\bar{q}_{0})]\rho^{1/2}(\bar{q}),\\
\bar{b}_{0}  &  =%
\begin{pmatrix}
\mu^{\prime}(q_{\text{eq}})\\
0
\end{pmatrix}
.
\end{align}
With these initial values, we propagate the time-dependent parameters $\bar
{q}_{t}$, $\bar{p}_{t}$, $\bar{A}_{t}$, and $\bar{\gamma}_{t}$ according to
Eqs.~(\ref{eq:q_t_dot})--(\ref{eq:gamma_t_dot}) and $\bar{b}_{t}$ according to
Eq.~(\ref{eq:b_t_dot}). The potential energy, its gradient, and its Hessian
are given by
\begin{align}
\bar{V}(\bar{q}_{t})  &  =V_{2}(q_{t})-V_{1}(q_{t}^{\prime}%
),\label{eq:V_q_bar}\\
\bar{V}^{\prime}(\bar{q}_{t})  &  =%
\begin{pmatrix}
V_{2}^{\prime}(q_{t})\\
-V_{1}^{\prime}(q_{t}^{\prime})
\end{pmatrix}
,\label{eq:grad_q_bar}\\
\bar{V}^{\prime\prime}(\bar{q}_{t})  &  =%
\begin{pmatrix}
V_{2}^{\prime\prime}(q_{t}) & 0\\
0 & -V_{1}^{\prime\prime}(q_{t}^{\prime})
\end{pmatrix}
, \label{eq:hess_q_bar}%
\end{align}
while the $D\times D$ mass matrix $m$ is replaced by the $2D\times2D$ matrix,
\begin{equation}
\bar{m}=%
\begin{pmatrix}
m & 0\\
0 & -m
\end{pmatrix}
, \label{eq:mass_q_bar}%
\end{equation}
where $q_{t}$ and $q_{t}^{\prime}$ are $D$-dimensional vectors composed of the first and second halves of coordinates of $\bar{q}_{t}$,
i.e., $\bar{q}_{t}=(q_{t},q_{t}^{\prime})^{T}$. Interestingly, the classical
equations of motion [Eqs.~(\ref{eq:q_t_dot}) and (\ref{eq:p_t_dot})] for the
parameters $\bar{q}_{t}$ and $\bar{p}_{t}$ are solved by propagating two
independent trajectories in $D$ spatial dimensions: the first trajectory
evolves $q_{t}$ and $p_{t}$ with the excited-state Hamiltonian $H_{2}$, while
the second trajectory evolves $q_{t}^{\prime}$ and $p_{t}^{\prime}$ with the
negative of the ground-state Hamiltonian, $-H_{1}$, due to the negative signs
of mass in Eq.~(\ref{eq:mass_q_bar}) and gradient in Eq.~(\ref{eq:grad_q_bar}%
). Because the second trajectory is at a fixed point, i.e., at the minimum of
the ground-state potential energy $V_{1}$ with zero momentum, it shows no
dynamics. As a result, one requires only a single excited-state classical
trajectory, to evolve the first $D$ coordinates of $\bar{q}_{t}$, and Hessians
of the excited-state potential energy surface along this trajectory, which is
the same as in the original zero-temperature approach; no further potential
energy evaluations are needed to account for the temperature effects.

Finally, let us note that an alternative approach to
finite-temperature spectra with Gaussian wavepackets has been proposed in
Ref.~\onlinecite{Reddy_Prasad:2016}. There, the authors propose a similar
scheme to directly evolve the coherence $\hat{\rho}_{\mu}(t)=\exp(-i\hat
{H}_{2}t/\hbar)\hat{\mu}\hat{\rho}\exp(i\hat{H}_{1}t/\hbar)$ in position
representation with the doubled number of degrees of freedom. Then, the
correlation function is evaluated simply as $C(t)=\text{Tr}[\hat{\mu}%
^{\dagger}\hat{\rho}_{\mu}(t)]$. Their method, combined with the local
harmonic approximation, is equivalent to ours and gives the same correlation
function. In contrast to our approach, the method of
Ref.~\onlinecite{Reddy_Prasad:2016} has so far been used to compute vibronic
spectra only in systems described with globally harmonic potential energy
surfaces, where it is equivalent to the global harmonic approximation for
vibronic spectra, which is much simpler because analytical expressions for
$C(t)$ exist.\cite{Baiardi_Barone:2013,Tapavicza_Sundholm:2016} Our approach
based on thermo-field dynamics has the advantage of reducing the transition
dipole autocorrelation function to the simpler and well-known expression
(\ref{eq:C_t_norep}) for the wavepacket autocorrelation, thus making it very
easy to implement the finite-temperature treatment of vibronic spectra into
the standard zero-temperature wavefunction-based codes, which typically
contain procedures for computing the wavepacket autocorrelation.

\section{\label{sec:compdet}Computational details}

\subsection{\label{subsec:morse_comp}Morse potential}

To test the accuracy of the proposed method, we construct a one-dimensional
model system consisting of a ground-state harmonic potential and an
excited-state Morse potential. The ground-state surface is assumed harmonic to
exclude the error (or error cancellation) due to using an approximate initial
vibrational state---this is rarely an issue with zero-temperature methods
because the harmonic approximation typically holds in the vicinity of the
potential minimum but could affect the results at higher temperatures. In the
current model, the error of the results obtained with thawed Gaussian
approximation is only due to the anharmonicity of the excited-state potential
energy surface.

A set of Morse potentials,
\begin{equation}
V_{2}(q)=V_{2}(q_{2})+\frac{\omega_{2}}{4\chi}[1-e^{-\sqrt{2m\omega_{2}\chi
}(q-q_{2})}]^{2}, \label{eq:Morse_pot}%
\end{equation}
was constructed by fixing the equilibrium position $q_{2}$, minimum energy
$V_{2}(q_{2})$, and frequency $\omega_{2}=\sqrt{V_{2}^{\prime\prime}(q_{2}%
)/m}$ at $q_{2}$ and by varying the anharmonicity parameter $\chi$. We set
the minimum of the ground-state harmonic potential to zero ($q_{1}=0$) and its
frequency to $\omega_{1}=1$, while the excited-state Morse parameters were
$q_{2}=1.5$, $\omega_{2}=0.9$, and $V(q_{2})=10$. Mass was set to $m=1$. The
level of anharmonicity was tuned by changing the parameter $\chi$ in the range
between $0.01$ and $0.02$, in steps of $0.001$.

The exact spectrum was computed by evaluating Franck--Condon
factors by numerical integration, which is feasible for this one-dimensional
model system since both harmonic and Morse vibrational eigenfunctions are
known analytically. The adiabatic harmonic model,
\begin{equation}
V_{2}^{\text{AH}}(q)=V_{2}(q_{2})+\frac{1}{2}m\omega_{2}^{2}(q-q_{2})^{2},
\label{eq:AH}%
\end{equation}
which is constructed about the minimum of the potential energy surface, is the
same for all constructed Morse potentials because it does not depend on $\chi
$. Since the (extended) thawed Gaussian approximation is exact for harmonic
potentials, it was used to compute the adiabatic harmonic spectra. For both
harmonic and thawed Gaussian dynamics calculations, time step was 0.1 and the
total simulation time was 1000, i.e., 10000 steps in total were taken.
Gaussian broadening with half-width at half-maximum of 0.1 was applied to all
spectra. Spectra were evaluated at scaled temperatures
$T_{\omega}=0$, $0.5$, and $1$, where $T_{\omega}=k_{\text{B}}T/\hbar
\omega_{1}=1/\beta\hbar\omega_{1}$ (e.g., for an average molecular
vibration of $\omega=1000\,\text{cm}^{-1}$, $T_{\omega}=1$ corresponds to the
temperature $T\approx1439\,$K). A constant transition dipole moment $\mu=1$
was used.

To compare reference ($\sigma_{\text{ref}}$) and approximate ($\sigma$)
spectra, we used the spectral contrast angle $\theta$, defined through its
cosine as
\begin{equation}
\cos\theta=\frac{\sigma_{\text{ref}}\cdot\sigma}{\lVert\sigma_{\text{ref}%
}\rVert\lVert\sigma\rVert}, \label{eq:cos_theta}%
\end{equation}
where $\sigma_{1}\cdot\sigma_{2}=\int d\omega\sigma_{1}(\omega)\sigma
_{2}(\omega)$ is the inner product of two spectra and $\lVert\sigma
\rVert=\sqrt{\sigma\cdot\sigma}$ is the associated norm. In all calculations, the
reference was the exact spectrum, while the approximate spectra were computed
with the adiabatic global harmonic and thawed Gaussian approximations.

\subsection{\label{subsec:otf_comp}On-the-fly \textit{ab initio} calculations}

The $\text{S}_{1} \leftarrow\text{S}_{0}$ absorption spectrum of benzene was
computed with adiabatic harmonic, vertical harmonic, and thawed Gaussian
approximations. In short, the adiabatic harmonic model is, as described above,
obtained by the second-order Taylor expansion of the excited-state potential
energy surface about its minimum, while for the vertical harmonic model, the
same expansion is performed about the ground-state minimum.

Density functional theory was used for the optimization and Hessian
calculation of the ground electronic state, while its time-dependent version
was employed for the excited-state optimization, energy, gradient, and Hessian
calculations. We used the B3LYP functional with the ultrafine grid
and 6-31+G(d,p) basis set, as implemented in the Gaussian09\cite{g09} package.
For the thawed Gaussian propagation, we used a second-order symplectic
integrator with a time step of 8 a.u. ($\approx0.2$\thinspace fs) and 10000
steps in total. The Hessian of the potential energy was evaluated every four
steps and interpolated in between, as done previously in
Ref.~\onlinecite{Patoz_Vanicek:2018}. The ground-state surface was assumed to
be harmonic. The gradient of the electronic transition dipole moment was
computed numerically by the second-order finite difference method with a step
of $10^{-4}$\thinspace\AA . \cite{Patoz_Vanicek:2018}

The computed correlation functions were multiplied by an exponential damping
function $e^{-t/\tau}$ with $\tau=18000$\thinspace a.u., resulting in a
Lorentzian line shape with half-width at half-maximum of $\approx
12.2$\thinspace cm$^{-1}$. To facilitate comparison with the experimental
spectrum, computed spectra were shifted and scaled to match the experimental
spectrum of Ref.~\onlinecite{Dawes_Mason:2017} at its highest peak (data taken
from the MPI-Mainz UV/VIS Spectral
Atlas\cite{spec_atlas:2020,Keller-Rudek_Sorensen:2013}).

Finally, let us emphasize that the finite-temperature treatment of spectra
requires no additional electronic structure evaluations, i.e., the same \textit{ab
initio} data could be reused to compute the benzene spectrum at any given
temperature. We evaluated the spectra at zero temperature and at the
temperature of the experiment ($T = 298$\,K).

\section{\label{sec:resanddisc}Results and discussion}

\subsection{\label{subsec:morse_res}Morse potential}

\begin{figure}
[t]%
\includegraphics[width=\textwidth]{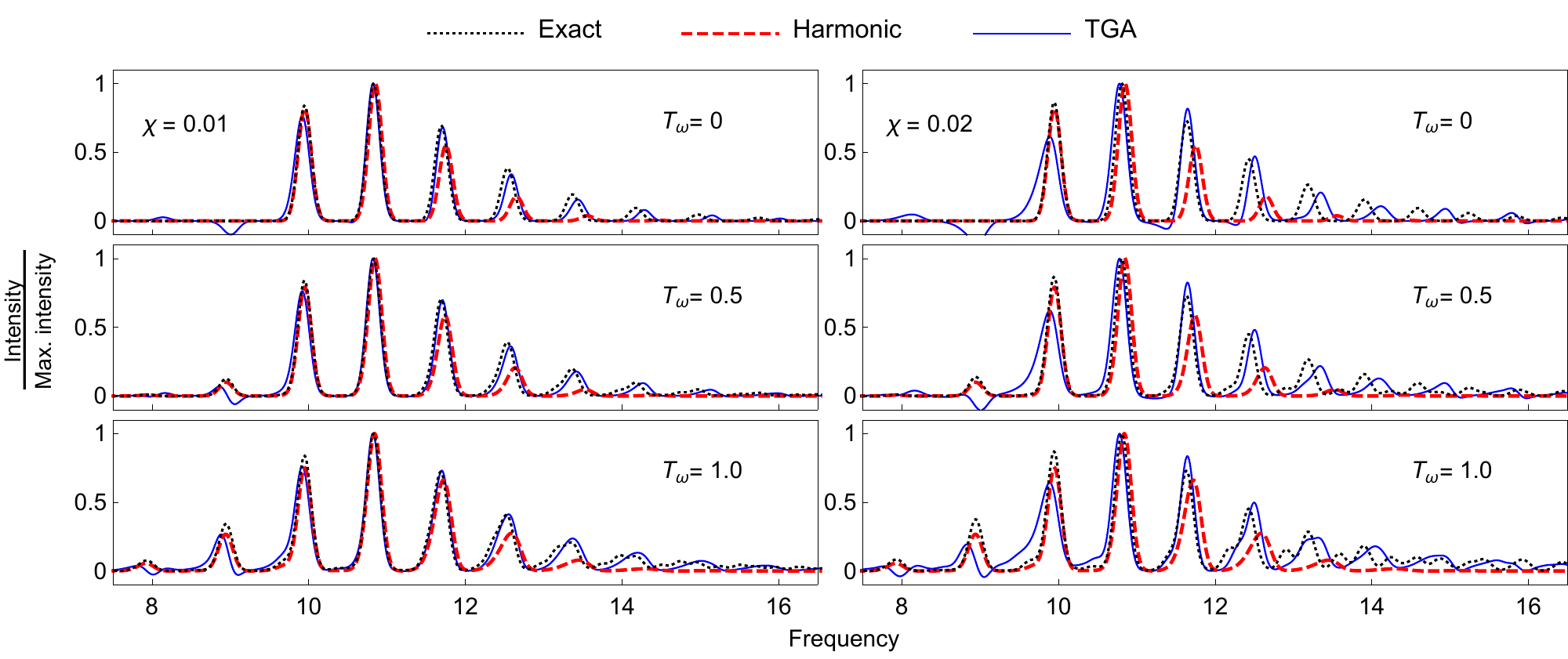}\caption{Exact,
thawed Gaussian (``TGA'', Sec.~\ref{subsec:spectraT}), and adiabatic harmonic
[Eq.~(\ref{eq:AH})] spectra for the Morse potential model systems (see
Sec.~\ref{subsec:morse_comp}) with lower (left panels) or higher (right
panels) degree of anharmonicity $\chi$ and at three different temperatures
$T_{\omega}=k_{\text{B}}T/\hbar\omega_{1}$ (see Sec.~\ref{subsec:morse_comp}).}\label{fig:Morse_Comparison_Spectra}%

\end{figure}

Thawed Gaussian and global harmonic spectra were compared with the exact
result (see Fig.~\ref{fig:Morse_Comparison_Spectra}). Already for the system
with weak anharmonicity (left panels, $\chi= 0.01$), the thawed Gaussian
approximation provides a more accurate spectrum than the harmonic method. The
difference is seen mainly in the intensities of the high-frequency peaks.
Since the adiabatic harmonic model describes well the region around the
potential minimum, it can recover the positions and intensities of peaks
corresponding to transitions between vibrational states with small quantum
numbers. In contrast, the harmonic approximation breaks down for vibrational
states with more quanta, resulting in incorrect intensities of high-frequency
transitions. The effect of anharmonicity on the peak positions becomes
significant for $\chi= 0.02$ and even the thawed Gaussian approximation is
inadequate. Nevertheless, it is still more accurate than the adiabatic
harmonic model, which has no dependence on $\chi$ (harmonic spectra are,
clearly, the same for different $\chi$ at a given temperature). In
typical molecular systems, the peaks are often left unresolved due to the
short excited-state lifetime or inhomogeneous broadening. Then, the
intensities play an important role in recovering the overall shape of the
spectrum, whereas even an error of tens of reciprocal centimeters in peak
positions can be tolerated.

In contrast to the global harmonic method, the thawed Gaussian approximation
can result in non-physical negative spectral features, which are
due to the nonlinear character of the Schr\"{o}dinger equation (\ref{eq:Sch}%
). This is a well-known disadvantage of the method and was
discussed in more detail
elsewhere.\cite{Wehrle_Vanicek:2015,Begusic_Vanicek:2019} In the studied
Morse system, a negative peak overlaps with the hot band around $\omega=9$,
resulting in poor description of this spectral region at higher temperatures.
In a way, the gain in accuracy in the high-frequency part of the spectrum is
accompanied by a loss in accuracy in the frequency region below the 0-0 transition.

\begin{figure}
[ht]\includegraphics[width=0.48\textwidth]{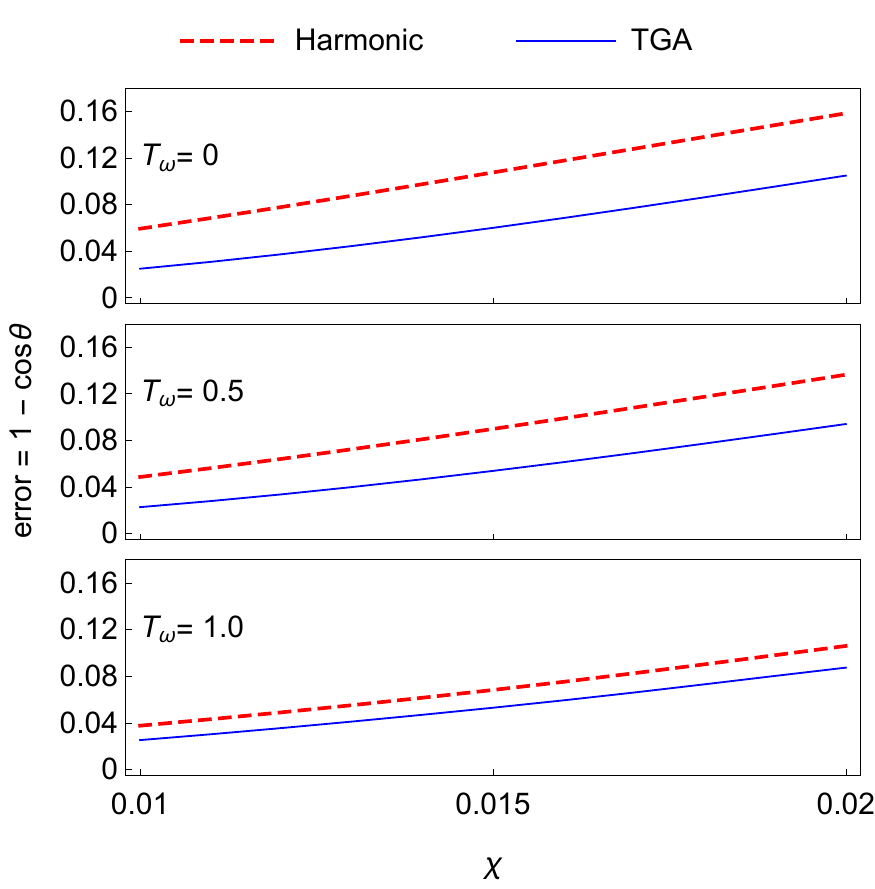} \caption{\label{fig:Morse_Comparison_Angles}Errors, measured by the spectral contrast angle [Eq.~(\ref{eq:cos_theta})], of the spectra computed with the thawed Gaussian approximation (``TGA'', Sec.~\ref{subsec:spectraT}) or the adiabatic harmonic approach [Eq.~(\ref{eq:AH})], as a function of the anharmonicity parameter $\chi$. Results are shown for three different temperatures $T_{\omega}=k_{\text{B}}T/\hbar\omega_{1}$ (see Sec.~\ref{subsec:morse_comp}).}
\end{figure}

To compare the global harmonic and thawed Gaussian methods quantitatively, we
measure the error of an approximate spectrum with the spectral contrast angle
between the approximate and exact spectra (see
Fig.~\ref{fig:Morse_Comparison_Angles}). The thawed Gaussian approximation
gives more accurate spectra than the harmonic approximation for all
anharmonicities and at all temperatures studied. However, an interesting trend
is observed: the harmonic approximation becomes more accurate as the
temperature increases, whereas the thawed Gaussian approximation keeps the
same degree of accuracy at all temperatures. The main reason for such behavior
is closely related to the discussion above. As the temperature increases, the
intensity of hot bands below the 0-0 transition grows and they become more
relevant in measuring the error. Hence, the adiabatic harmonic method gains on
accuracy, unlike the thawed Gaussian approximation, which always loses on
accuracy in the low-frequency part of the spectrum. However, such behavior of
the global harmonic method is not general; if the ground-state potential
energy surface were anharmonic, high-temperature spectra would also reflect
the effects neglected in the global harmonic models---those of ground-state
anharmonicity on the initial density matrix.

\subsection{\label{subsec:otf_res}Absorption spectrum of benzene}

The symmetry-forbidden S$_{1} \leftarrow$ S$_{0}$ transition in benzene is a
well-known example of the Herzberg--Teller
effect,\cite{Herzberg:1966,Hollas:2004} where the spectrum arises only due to
the coordinate dependence of the transition dipole moment, which is zero by
symmetry at the equilibrium geometry. As such, it has been studied extensively
both from the
experimental\cite{Atkinson_Parmenter:1978,Trost_Platt:1997,Etzkorn:1999,Loginov_Drabbels:2008,Fally_Vandaele:2009,Dawes_Mason:2017}
and
theoretical\cite{Sponer_Teller:1939,Faulkner_Richardson:1979,Fischer_Knight:1992,He_Pollak:2001,Worth:2007,Penfold_Worth:2009,Li_Lin:2010,Crespo-Otero_Barbatti:2012}
points of view. The spectrum is a challenge for computational methods because
it is highly resolved, exhibits Herzberg--Teller effects, and contains hot
bands due to finite temperature. Although benzene is typically considered to
be a rigid molecule, we have recently shown that the anharmonicity affects
significantly the intensities of the peaks in the main progression of the
spectrum.\cite{Patoz_Vanicek:2018,Begusic_Vanicek:2018} However, our previous
work assumed zero temperature, therefore neglecting the weak hot bands present
in the experimental spectrum.

\begin{figure}
[th]\includegraphics[width = \textwidth]{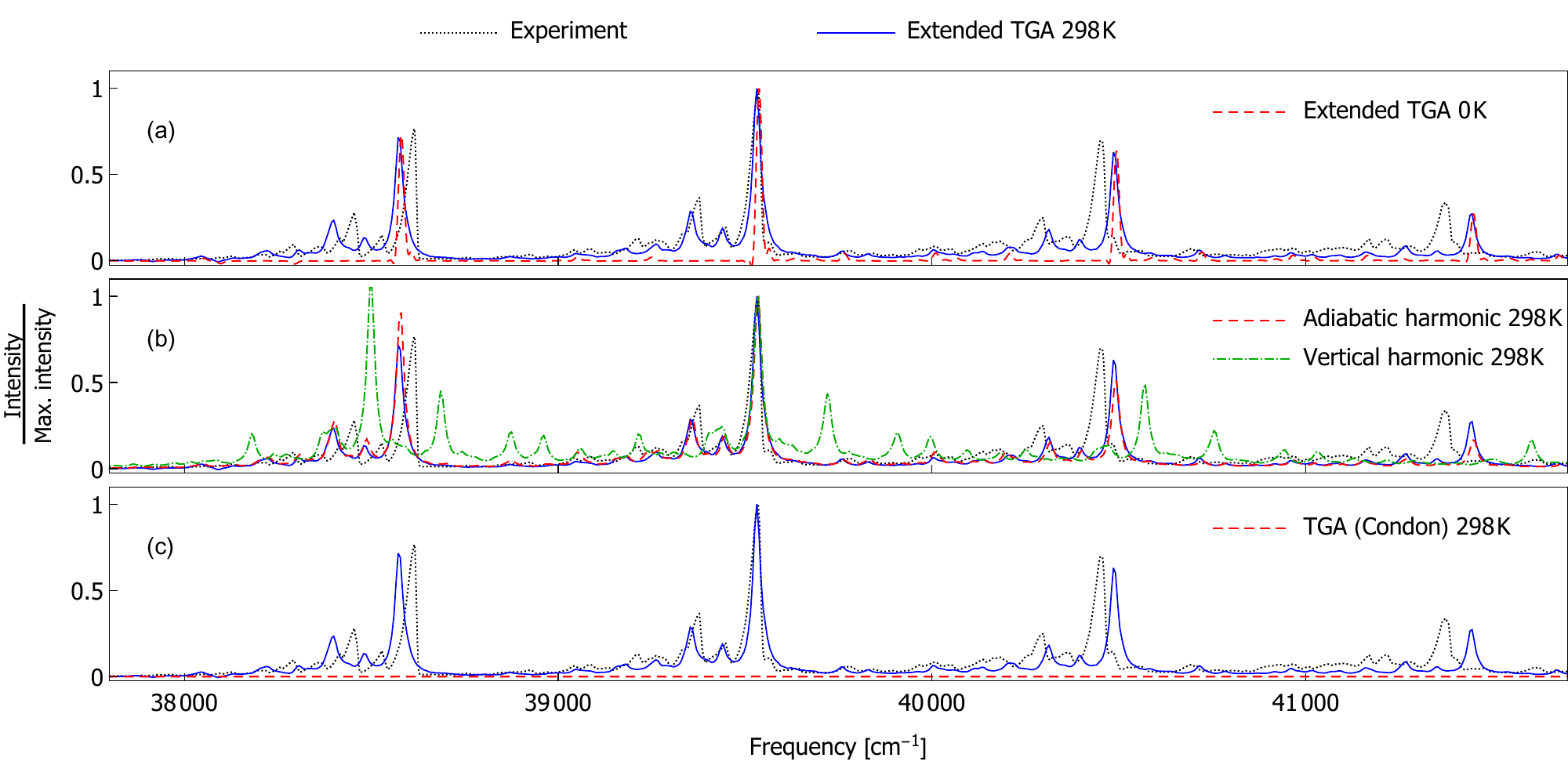}\caption{Benzene S$_{1}
\leftarrow$ S$_{0}$ absorption spectrum computed with the extended thawed
Gaussian approximation (``Extended TGA'') at 298\,K (using the approach
described in Sec.~\ref{subsec:tga_T}), compared with the experimental
spectrum\cite{Dawes_Mason:2017,Keller-Rudek_Sorensen:2013} measured at 298\,K
and other approximate spectra simulations based on: (a) zero-temperature
extended thawed Gaussian approximation (``Extended TGA 0\,K'') as described in
Sec.~\ref{subsec:tga_0}, (b) adiabatic or vertical global harmonic models at
298\,K (see Sec.~\ref{subsec:otf_comp}), and (c) thawed Gaussian
approximation, which assumes Condon approximation [``TGA (Condon)''].}\label{fig:Benzene_All}%

\end{figure}

Here, we complement our earlier result with the new finite-temperature
extended thawed Gaussian method. First, we demonstrate
[Fig.~\ref{fig:Benzene_All}(a)] the effect of non-zero temperature on the
spectrum. Whereas the original, zero-temperature extended thawed Gaussian
approximation neglects completely the weak, but non-negligible, hot bands, the
finite-temperature approach reproduces all features of the spectrum. The
inaccuracy in the frequencies of the peaks is most likely due to the
electronic structure method used; we discuss this later. Nevertheless,
Fig.~\ref{fig:Benzene_All}(a) clearly shows the difference in the spectra
computed without and with finite-temperature effects.

We argue that the benzene absorption spectrum is affected by the anharmonicity
of the excited-state potential energy surface. This effect is best
demonstrated by the difference in spectra based on two global harmonic models:
if the potential energy surface were harmonic, the second-order expansion of
the potential energy about any molecular geometry would result in the same
spectrum. As shown in Fig.~\ref{fig:Benzene_All}(b), in benzene, the adiabatic
harmonic method is much more accurate than the vertical; in general, either of
the two methods can be more
appropriate.\cite{Wehrle_Vanicek:2014,Wehrle_Vanicek:2015,Hazra_Nooijen:2005}
The extended thawed Gaussian approximation outperforms not only the vertical
harmonic approach, whose spectrum is completely off, but also the adiabatic
harmonic approximation, which fails to produce accurate peak intensities.

Figure~\ref{fig:Benzene_All}(c) shows the importance of treating the
Herzberg--Teller effect with the extended thawed Gaussian approximation. Since
the transition is symmetry-forbidden, i.e., $\mu(q_{\text{eq}}) = 0$, the
spectrum computed within the Condon approximation [$\mu(q) \approx
\mu(q_{\text{eq}})$] vanishes, whereas the full, Herzberg--Teller treatment
reproduces the experimental spectrum.

\begin{figure}
[th]%
\includegraphics[width = \textwidth]{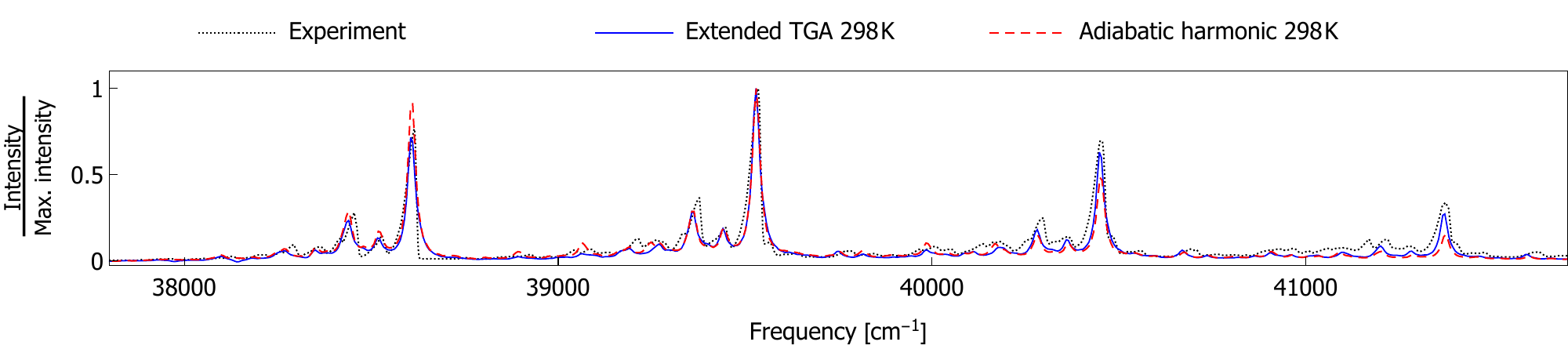}\caption{Benzene
S$_{1} \leftarrow$ S$_{0}$ absorption spectra computed with the extended
thawed Gaussian approximation (``Extended TGA'') and adiabatic harmonic model,
both at 298\,K, compared with the experimental
spectrum\cite{Dawes_Mason:2017,Keller-Rudek_Sorensen:2013} measured at 298\,K.
The adiabatic harmonic model was modified by scaling both ground- and
excited-state frequencies by a constant $f = 0.963$, which was taken from
Ref.~\onlinecite{cccbdb_2019} and is associated with the electronic structure
method used (see Sec.~\ref{subsec:otf_comp}). For the spectrum evaluated with
the extended thawed Gaussian approximation, we applied the same scaling factor
only to the values on the frequency axis.}\label{fig:Benzene_Anharmonicity_Scaled}%

\end{figure}

In computational chemistry, vibrational scaling factors,\cite{cccbdb_2019}
which we denote by $f$, are often used to empirically correct for systematic
errors in the vibrational frequencies computed with electronic structure
methods. In vibronic spectroscopy, such scaling, applied to ground- and
excited-state frequencies, can modify both peak positions and
intensities.\cite{Baiardi_Barone:2013} However, the effect on intensities is
often weak; indeed, the adiabatic harmonic spectrum with scaled vibrational
frequencies (red, dashed line in Fig.~\ref{fig:Benzene_Anharmonicity_Scaled})
exhibits almost perfect peak positions, but still the same errors in
intensities as the adiabatic harmonic spectrum of Fig.~\ref{fig:Benzene_All}%
(b). For comparison---and for comparison only---we show an analogous,
\textquotedblleft corrected\textquotedblright\ spectrum computed with the
extended thawed Gaussian approximation (blue, solid line in
Fig.~\ref{fig:Benzene_Anharmonicity_Scaled}). Since the simple procedure of
scaling the vibrational frequencies is not applicable in this case, we scale
directly the frequency axis by $f$, which corrects peak positions but leaves
intensities unchanged. The results imply that the subtle anharmonicity effects
on spectral intensities, described well with the on-the-fly semiclassical
thawed Gaussian method, cannot be captured even with the corrected harmonic potential.

\section{Conclusion}

In conclusion, we have presented a new approach to compute vibronic spectra at
finite temperature within the framework of the thawed Gaussian approximation.
The proposed method describes partially the effect of anharmonicity on the
spectrum and at the same time includes all effects treated in the conventional
global harmonic approach---mode-mode coupling, non-zero temperature, and
Herzberg--Teller contribution to the transition dipole moment. Most
importantly, the inclusion of finite temperature comes at no additional
computational cost or deterioration in accuracy. Hence, the proposed procedure
provides a viable route to systematically improve on global harmonic
simulations at any temperature. Finally, this on-the-fly \textit{ab initio}
semiclassical approach to thermo-field dynamics could inspire other quantum or
semiclassical \textquotedblleft direct dynamics\textquotedblright\ methods for
computing spectra at finite temperatures.

The data that support the findings of this study are available from the
corresponding author upon reasonable request.

\begin{acknowledgments}
The authors would like to thank Maxim Gelin for introducing them to
thermo-field dynamics and Lipeng Chen for helpful discussions. The financial
support from the European Research Council (ERC) under the European Union's
Horizon 2020 research and innovation programme (grant agreement No. 683069 --
MOLEQULE) is gratefully acknowledged.
\end{acknowledgments}

\appendix

\section{Relation to thermo-field dynamics \label{app_sec:TFD}}

Following Suzuki,\cite{Suzuki:1985} let us define
\begin{equation}
| \bar{I} \rangle= \sum_{k} | k \tilde{k} \rangle= | 0 \tilde{0} \rangle+ | 1
\tilde{1} \rangle+ \ldots,
\end{equation}
where $| k \tilde{k} \rangle$ denotes a basis vector of a space obtained as a
direct product of ``physical'' (with basis $\lbrace|k\rangle\rbrace$) and
``fictitious'' (with basis $\lbrace|\tilde{k}\rangle\rbrace$) Hilbert spaces.
In general, we use tilde $\ \tilde{ }\ $ to denote an element of (or an
operator acting on) the ``fictitious'' Hilbert space and bar $\ \bar{ }\ $ (as
opposed to bold font used in Ref.~\onlinecite{Borrelli_Gelin:2016}) for the
direct-product space. The physical and fictitious states are related through
the conjugation rule\cite{Suzuki:1985}
\begin{equation}
(u_{1} | k \rangle+ u_{2} | k^{\prime} \rangle)\ \tilde{ } = u^{\ast}_{1} |
\tilde{k} \rangle+ u^{\ast}_{2} | \tilde{k}^{\prime} \rangle,
\label{eq:conjg_rule_1}%
\end{equation}
which results in
\begin{equation}
\langle\tilde{\alpha} | \hat{\tilde{A}} | \tilde{\alpha}^{\prime} \rangle=
\langle\alpha^{\prime} | \hat{A} | \alpha\rangle\label{eq:conjg_rule_2}%
\end{equation}
for arbitrary complex numbers $u_{1}$ and $u_{2}$, states $|\alpha\rangle$ and
$|\alpha^{\prime}\rangle$, and operator $\hat{A}$. Next, the so-called thermal
vacuum is defined as
\begin{equation}
| \bar{0}(\beta) \rangle= \hat{\rho}^{1/2} | \bar{I} \rangle,
\end{equation}
where $\hat{\rho}$ is the density operator and acts only on the physical
Hilbert space. Then, the correlation function, defined in Eq.~(\ref{eq:C_t_1}%
), can be written as
\begin{equation}
C(t) = \langle\bar{\phi}_{0} | e^{-i\hat{\bar{H}}t/\hbar} | \bar{\phi}_{0}
\rangle, \label{eq:C_t_TFD}%
\end{equation}
where $| \bar{\phi}_{0} \rangle= \hat{\mu} | \bar{0}(\beta) \rangle$ and
\begin{equation}
\hat{\bar{H}} = \hat{H}_{2} - \hat{\tilde{H}}_{1}. \label{eq:H_bar_TFD}%
\end{equation}
The proof goes as follows:
\begin{align}
C(t)  &  = \langle\bar{0}(\beta) | \hat{\mu}^{\dagger} e^{-i\hat{\bar{H}%
}t/\hbar} \hat{\mu} | \bar{0}(\beta) \rangle\label{eq:C_t_TFD_1}\\
&  = \sum_{k,k^{\prime}} \langle k \tilde{k} | \hat{\rho}^{1/2} \hat{\mu
}^{\dagger} e^{-i\hat{\bar{H}} t /\hbar} \hat{\mu} \hat{\rho}^{1/2} |
k^{\prime} \tilde{k}^{\prime} \rangle\label{eq:C_t_TFD_2}\\
&  = \sum_{k,k^{\prime}} \langle k | \hat{\rho}^{1/2} \hat{\mu}^{\dagger}
e^{-i\hat{H}_{2} t /\hbar} \hat{\mu} \hat{\rho}^{1/2} | k^{\prime}
\rangle\langle\tilde{k} | e^{i\hat{\tilde{H}}_{1} t /\hbar} | \tilde
{k}^{\prime} \rangle\label{eq:C_t_TFD_3}\\
&  = \sum_{k,k^{\prime}} \langle k | \hat{\rho}^{1/2} \hat{\mu}^{\dagger}
e^{-i\hat{H}_{2} t /\hbar} \hat{\mu} \hat{\rho}^{1/2} | k^{\prime}
\rangle\langle k^{\prime} | e^{i\hat{H}_{1} t /\hbar} | k \rangle
\label{eq:C_t_TFD_4}\\
&  = \sum_{k} \langle k | \hat{\rho}^{1/2} \hat{\mu}^{\dagger} e^{-i\hat
{H}_{2} t /\hbar} \hat{\mu} \hat{\rho}^{1/2} e^{i\hat{H}_{1} t /\hbar} | k
\rangle\label{eq:C_t_TFD_5}\\
&  = \text{Tr} ( \hat{\mu}^{\dagger} e^{-i\hat{H}_{2} t /\hbar} \hat{\mu}
\hat{\rho} e^{i\hat{H}_{1} t /\hbar}), \label{eq:C_t_TFD_6}%
\end{align}
where we used (\ref{eq:conjg_rule_2}) to go from (\ref{eq:C_t_TFD_3}) to
(\ref{eq:C_t_TFD_4}).

To complete the equivalence between Eq.~(\ref{eq:C_t_TFD}) and
Eq.~(\ref{eq:C_t_norep}) of the main text, we demonstrate that
\begin{equation}
\langle q \tilde{q}^{\prime} | \bar{\phi}_{0} \rangle= \bar{\phi}%
_{0}(q,q^{\prime}),
\end{equation}
i.e., that the position representation of state $| \bar{\phi}_{0} \rangle$
introduced in this appendix is the function defined in Eq.~(\ref{eq:phi_0_bar}%
) of the main text. Indeed,
\begin{align}
\langle q \tilde{q}^{\prime} | \bar{\phi}_{0} \rangle &  = \sum_{k} \langle q
| \hat{\mu} \hat{\rho}^{1/2} | k \rangle\langle\tilde{q}^{\prime} | \tilde{k}
\rangle\\
&  = \sum_{k} \langle q | \hat{\mu} \hat{\rho}^{1/2} | k \rangle\langle k |
q^{\prime}\rangle\\
&  = \langle q | \hat{\mu} \hat{\rho}^{1/2} | q^{\prime}\rangle\\
&  = \bar{\phi}_{0}(q,q^{\prime}),
\end{align}
where we again used Eq.~(\ref{eq:conjg_rule_2}) with $\hat{A}$ being identity operator.

\section{Derivation of the initial-state parameters
\label{app_sec:density}}

Here, we derive expressions~(\ref{eq:A_0_temp})--(\ref{eq:gamma_0_temp}) for
the Gaussian parameters of $\rho^{1/2}(q,q^{\prime})$. First, recall that the
matrix element $\rho(q,q^{\prime})$ of the thermal density operator in a
one-dimensional harmonic oscillator
\begin{equation}
\hat{H}=\frac{\hat{p}^{2}}{2m}+\frac{1}{2}m\omega^{2}\hat{q}^{2}%
\end{equation}
with mass $m$ and frequency $\omega$ is\cite{feynman:1965}
\begin{align}
\rho(q,q^{\prime}) &  =\sqrt{\frac{m\omega\tanh(\beta\hbar\omega/2)}{\pi\hbar
}}\nonumber\\
&  \times e^{-\frac{m\omega}{2\hbar}[(q^{2}+q^{\prime2})\coth(\beta\hbar
\omega)-2qq^{\prime}/\sinh(\beta\hbar\omega)]}.\label{eq:rho_1D}%
\end{align}
Using Eq.~(\ref{eq:rho_1D}), we now derive the expression for the matrix
element $\rho(q,q^{\prime})$ in a $D$-dimensional coupled harmonic oscillator
\begin{equation}
\hat{H}=\frac{1}{2}\hat{p}^{T}\cdot m^{-1}\cdot\hat{p}+\frac{1}{2}(\hat
{q}-q_{\text{eq}})^{T}\cdot K\cdot(\hat{q}-q_{\text{eq}}),
\end{equation}
where $m$ and $K$ are $D\times D$ symmetric mass and force-constant matrices,
respectively, and $q_{\text{eq}}$ is the equilibrium position at which the
potential energy has its minimum.

The first step consists in transforming $q$ to the mass-scaled normal-mode
coordinates $Q=O^{T}\cdot m^{1/2}\cdot(q-q_{\text{eq}})$, where $O$ is the
orthogonal matrix diagonalizing the mass-scaled force-constant matrix, i.e.,
$O^{T}\cdot m^{-1/2}\cdot K\cdot m^{-1/2}\cdot O=\Omega_{\text{diag}}^{2}$
with a real diagonal matrix $\Omega_{\text{diag}}$. This leads to the
uncoupled Hamiltonian
\begin{equation}
\hat{H}=\frac{1}{2}\hat{P}^{T}\cdot\hat{P}+\frac{1}{2}\hat{Q}^{T}\cdot
\Omega_{\text{diag}}^{2}\cdot\hat{Q}=\sum_{i=1}^{D}\hat{H}_{i},
\end{equation}
where $\hat{H}_{i}=\frac{1}{2}\hat{P}_{i}^{2}+\frac{1}{2}\omega_{i}^{2}%
\hat{Q_{i}}^{2}$ and $\omega_{i}$ are the diagonal elements of $\Omega
_{\text{diag}}$. The density matrix element of this uncoupled $D$-dimensional
harmonic oscillator is
\begin{align}
\rho(Q,Q^{\prime})  &  =\frac{\langle Q|e^{-\beta\hat{H}}|Q^{\prime}\rangle
}{\operatorname{Tr}(e^{-\beta\hat{H}})}\\
&  =\prod_{i=1}^{D}\frac{\langle Q_{i}|e^{-\beta\hat{H}_{i}}|Q_{i}^{\prime
}\rangle}{\operatorname{Tr}(e^{-\beta\hat{H}_{i}})}\\
&  =\prod_{i=1}^{D}\rho_{i}(Q_{i},Q_{i}^{\prime})\\
&  =\prod_{i=1}^{D}\sqrt{\frac{\omega_{i}\tanh(\beta\hbar\omega_{i}/2)}%
{\pi\hbar}}\nonumber\\
&  \times e^{-\omega_{i}[(Q_{i}^{2}+Q_{i}^{\prime\,2})\coth(\beta\hbar
\omega_{i})-2Q_{i}Q_{i}^{\prime}/\sinh(\beta\hbar\omega_{i})]/2\hbar}\\
&  =\sqrt{\prod_{i=1}^{D}\left(  \frac{\omega_{i}\tanh(\beta\hbar\omega
_{i}/2)}{\pi\hbar}\right)  }\nonumber\\
&  \times e^{-\frac{1}{2\hbar}\sum_{i=1}^{D}[\omega_{i}(Q_{i}^{2}%
+Q_{i}^{\prime\,2})\coth(\beta\hbar\omega_{i})-\frac{2\omega_{i}}{\sinh
(\beta\hbar\omega_{i})}Q_{i}Q_{i}^{\prime}]}\\
&  =\sqrt{\frac{\det[\Omega_{\text{diag}}\cdot\tanh(\beta\hbar\Omega
_{\text{diag}}/2)]}{\pi^{D}\hbar^{D}}}\nonumber\\
&  \times\exp\left\{  \frac{i}{2\hbar}%
\begin{pmatrix}
Q, & Q^{\prime}%
\end{pmatrix}
\cdot\bar{A}_{\beta,\text{diag}}\cdot%
\begin{pmatrix}
Q\\
Q^{\prime}%
\end{pmatrix}
\right\}  ,
\end{align}
where
\begin{align}
\bar{A}_{\beta,\text{diag}}  &  =i\bar{\Omega}_{\text{diag}}\cdot%
\begin{pmatrix}
\coth(\beta\hbar\Omega_{\text{diag}}) & -\sinh(\beta\hbar\Omega_{\text{diag}%
})^{-1}\\
-\sinh(\beta\hbar\Omega_{\text{diag}})^{-1} & \coth(\beta\hbar\Omega
_{\text{diag}})
\end{pmatrix}
,\\
\bar{\Omega}_{\text{diag}}  &  =%
\begin{pmatrix}
\Omega_{\text{diag}} & 0\\
0 & \Omega_{\text{diag}}%
\end{pmatrix}
\end{align}
are $2D\times2D$ matrices composed of $D\times D$ diagonal sub-matrices.

Transformation back to the original coordinates $q$ yields
\begin{align}
\rho(q,q^{\prime})  &  =\sqrt{\frac{\det[m\cdot\Omega\cdot\tanh(\beta
\hbar\Omega/2)]}{\pi^{D}\hbar^{D}}}\nonumber\\
&  \times\exp\left[  \frac{i}{2\hbar}(\bar{q}-\bar{q}_{0})^{T}\cdot\bar
{A}_{\beta}\cdot(\bar{q}-\bar{q}_{0})\right]  ,
\end{align}
with
\begin{align}
\bar{q}  &  =%
\begin{pmatrix}
q\\
q^{\prime}%
\end{pmatrix}
,\qquad\bar{q}_{0}=%
\begin{pmatrix}
q_{\text{eq}}\\
q_{\text{eq}}%
\end{pmatrix}
,\\
\Omega &  =O\cdot\Omega_{\text{diag}}\cdot O^{T}=\sqrt{m^{-1/2}\cdot K\cdot
m^{-1/2}},\label{eq:Omega}\\
\bar{A}_{\beta}  &  =\bar{L}\cdot\bar{A}_{\beta,\text{diag}}\cdot\bar{L}^{T}=i%
\begin{pmatrix}
\text{A}_{\beta} & \text{B}_{\beta}\\
\text{B}_{\beta} & \text{A}_{\beta}%
\end{pmatrix}
,\label{eq:A_bar_beta}\\
\bar{L}  &  =%
\begin{pmatrix}
m^{1/2}\cdot O & 0\\
0 & m^{1/2}\cdot O
\end{pmatrix}
,\\
\text{A}_{\beta}  &  =m^{1/2}\cdot\Omega\cdot\coth(\beta\hbar\Omega)\cdot
m^{1/2},\label{eq:A_beta}\\
\text{B}_{\beta}  &  =-m^{1/2}\cdot\Omega\cdot\sinh(\beta\hbar\Omega
)^{-1}\cdot m^{1/2}. \label{eq:B_beta}%
\end{align}

To find $\rho^{1/2}(q,q^{\prime})$, we rewrite it as
\begin{equation}
\rho_{\beta}^{1/2}(q,q^{\prime})=\frac{\operatorname{Tr}(e^{-\beta\hat{H}/2}%
)}{\sqrt{\operatorname{Tr}(e^{-\beta\hat{H}})}}\rho_{\beta/2}(q,q^{\prime}),
\label{eq:SQRT_of_rho}%
\end{equation}
and hence, the initial value $\bar{A}_{0}$ needed for extended thawed Gaussian
propagation is given by $\bar{A}_{\beta/2}$, where $\bar{A}_{\beta}$ is
defined by Eqs.~(\ref{eq:A_bar_beta}), (\ref{eq:A_beta}), and (\ref{eq:B_beta}%
), which proves Eqs.~(\ref{eq:A_0_temp})--(\ref{eq:A_0_temp_B}). To find the
initial value $\bar{\gamma}_{0}$, we can avoid computing the traces in
Eq.~(\ref{eq:SQRT_of_rho}) explicitly and instead recognize that $\rho
^{1/2}(q,q^{\prime})$ must be normalized as
\begin{equation}
\int dqdq^{\prime}[\rho^{1/2}(q,q^{\prime})]^{2}=\text{Tr}[(\rho^{1/2}%
)^{2}]=\text{Tr}(\rho)=1.
\end{equation}
Therefore, $\bar{\gamma}_{0}$ can be computed from $\bar{A}_{0}$ in analogy
to Eq.~(\ref{eq:gamma_0}) for the wavepacket (\ref{eq:tga_wp}), i.e.,
\begin{align}
\bar{\gamma}_{0}  &  =-(i\hbar/4)\ln[\det(\text{Im}\bar{A}_{0}/\pi\hbar)]\\
&  =-(i\hbar/2)\ln[\det(m\cdot\Omega/\pi\hbar)]\\
&  =-(i\hbar/4)\ln[\det(m\cdot K/(\pi\hbar)^{2})],
\end{align}
where we used that the initial width matrix $\bar{A}_{0}$ is $\bar{A}%
_{\beta/2}$ and that
\[
\det(\text{Im}\bar{A}_{\beta})=\det(m\cdot\Omega)^{2}=\det(m\cdot K)
\]
(in particular, the determinant is independent of $\hbar$ and temperature!)
because%
\begin{align}
&  \det(\text{Im}\bar{A}_{\beta})\\
&  =\det(\text{A}_{\beta})\det(\text{B}_{\beta})\nonumber\\
&  \times\det(\text{A}_{\beta}\cdot\text{B}_{\beta}^{-1}-\text{B}_{\beta}%
\cdot\text{A}_{\beta}^{-1})\\
&  =\det(m\cdot\Omega)^{2}\nonumber\\
&  \times\det[\coth(\beta\hbar\Omega)]\det[-\sinh(\beta\hbar\Omega
)]^{-1}\nonumber\\
&  \times\det[-\sinh(\beta\hbar\Omega)^{2}\cdot\cosh(\beta\hbar\Omega)^{-1}]\\
&  =\det(m\cdot\Omega)^{2}=\det(m\cdot K).
\end{align}
{In this derivation, we used the relation\cite{Petersen_Pedersen:2012}
\begin{equation}
\det%
\begin{pmatrix}
\text{A} & \text{B}\\
\text{B} & \text{A}%
\end{pmatrix}
=\det(\text{A})\det(\text{B})\det(\text{A}\cdot\text{B}^{-1}-\text{B}%
\cdot\text{A}^{-1})
\end{equation}
valid for arbitrary invertible matrices $\text{A}$ and $\text{B{, }}%
${the relation}}
\begin{align*}
&  \text{A}_{\beta}\cdot\text{B}_{\beta}^{-1}-\text{B}_{\beta}\cdot
\text{A}_{\beta}^{-1}\\
&  =-m^{1/2}\cdot\Omega\cdot\left[  \cosh(\beta\hbar\Omega)-\cosh(\beta
\hbar\Omega)^{-1}\right]  \cdot\Omega^{-1}\cdot m^{-1/2}\\
&  =-m^{1/2}\cdot\Omega\cdot\sinh(\beta\hbar\Omega)^{2}\cosh(\beta\hbar
\Omega)^{-1}\cdot\Omega^{-1}\cdot m^{-1/2}%
\end{align*}
satisfied by matrices A$_{\beta}$ and B$_{\beta}$ from Eqs.~(\ref{eq:A_beta})
and (\ref{eq:B_beta}), and the definition (\ref{eq:Omega}) of $\Omega$ in the
last step.

\bibliographystyle{aipnum4-2}
\bibliography{biblio47, additions_FiniteTemperature}

\end{document}